\documentclass{aa}
\usepackage{psfig}
\usepackage{graphicx}

\def\bron{IGR~J17544-2619}
\def\optical{2MASS J17542527-2619526}
\def\ecs{erg~cm$^{-2}$s$^{-1}$}
\def\lum{erg~s$^{-1}$}

\begin{document}

\title{Chandra observation of the fast X-ray transient \bron:
evidence for a neutron star?}

\titlerunning{Chandra observation of \bron} 
\authorrunning{J.J.M. in 't Zand}

\author{
J.J.M.~in~'t~Zand\inst{1,2}
}


\institute{     SRON National Institute for Space Research, Sorbonnelaan 2,
                NL - 3584 CA Utrecht, the Netherlands 
	 \and
                Astronomical Institute, Utrecht University, P.O. Box 80000,
                NL - 3508 TA Utrecht, the Netherlands
	}

\date{Draft; version \today}

\abstract{\bron\ belongs to a distinct group of at least seven fast
X-ray transients that cannot readily be associated with nearby flare
stars or pre-main sequence stars and most probably are X-ray binaries
with wind accretion. Sofar, the nature of the accretor has been
determined in only one case (SAX J1819.3-2525/V4641~Sgr). We carried
out a 20~ks Chandra ACIS-S observation of \bron\ which shows the
source in quiescence going into outburst. The Chandra position
confirms the previous tentative identification of the optical
counterpart, a blue O9Ib supergiant at 3 to 4 kpc (Pellizza, Chaty \&
Negueruela, in prep.). This is the first detection of a fast X-ray
transient in quiescence. The quiescent spectrum is very soft. The
photon index of 5.9$\pm1.2$ (90\% confidence error margin) is much
softer than 6 quiescent black hole candidates that were observed with
Chandra ACIS-S (Kong et al. 2002; Tomsick et al.  2003). Assuming that
a significant fraction of the quiescent photons comes from the
accretor and not the donor star, we infer that the accretor probably
is a neutron star.  A fit to the quiescent spectrum of the neutron
star atmosphere model developed by Pavlov et al. (1992) and Zavlin et
al. (1996) implies an unabsorbed quiescent 0.5--10 keV luminosity of
$(5.2\pm1.3)\times10^{32}$~\lum. We speculate on the nature of the
brief outbursts. \keywords{X-rays: binaries -- X-rays: transients --
X-rays: individual: \bron} }

\maketitle 

\section{Introduction}
\label{intro}

It has recently become clear that there is a distinct group of fast
X-ray transients with outburst durations of order a few hours and peak
fluxes of order 10$^{-9}$~\ecs\ (2--10 keV) that are not readily
identifiable with other similarly active X-ray sources: magnetically
active nearby stars (i.e., DY Dra, RS CVn or pre-main sequence stars)
or superbursters (e.g., Kuulkers 2004; time profiles, luminosities and
spectra are inconsistent). The high fluxes and the lack of nearby
counterparts suggest high luminosities which would indicate an X-ray
binary origin. The first case, XTE~J1739--302, was discovered by Smith
et al. (1998). Accurate Chandra localization yielded an identification
with a highly reddened probably supergiant O star (Smith et
al. 2003a). The list has been steadily growing and now consists of at
least seven objects: XTE J1739--302, SAX J1818.6--1703 (In 't Zand et
al. 1998), V4641 Sgr (In 't Zand et al. 2000; Orosz et al. 2001), XTE
J1901+014 (Remillard \& Smith 2002), AX~J1749.1--2733 (Grebenev
2004b), IGR J16465-4507 (Lutovinov et al. 2005; Neguerela et al. 2005)
and \bron\ which is the subject of this paper. For the first two
systems there is suggestive evidence that the strong variability is
due to a high-mass companion star feeding the compact object through a
wind instead of an accretion disk, like in many high-mass X-ray
binaries (Smith 2004).

\bron\ was first reported from INTEGRAL detections on Sep. 17, 2003,
when it exhibited two flares (Sunyaev et al. 2003; Grebenev et
al. 2003) and a second time on Mar. 8, 2004 (Grebenev et
al. 2004a). Three XMM-Newton observations (Gonzalez-Riestra et
al. 2003, 2004) show that even outside large flares, the flux appears
to vary violently, with fluxes between an upper limit of
5$\times10^{-14}$~\ecs\ and 4$\times10^{-11}$~\ecs\ (0.5--10 keV). The
first detection of \bron\ was with the BeppoSAX Wide Field Cameras
that detected hours-long flares in 1996, 1999 and 2000 with 0.5--10
keV peak fluxes of a few times 10$^{-9}$~\ecs\ (In 't Zand et
al. 2004).  Semi-weekly flux measurements since 1999 with the RXTE
Proportional Counter Array (Swank \& Markwardt 2001) point to a duty
cycle of about 5\% above $\sim10^{-11}$~\ecs (2--10 keV; In 't Zand et
al. 2004).

Rodriguez (2003) tentatively identified a counterpart in the 2MASS
catalog, \optical, which is also in the USNO B1.0 catalog. Its
brightness is $B2=13.9$, $R2=12.0$ and $K_{\rm s}=8.02$. Pellizza,
Chaty \& Negueruela (in prep.) identify the optical counterpart as a
blue supergiant of spectral type O9Ib at 3 to 4 kpc distance. The
nature of the accretor is thus far undetermined. Rodriguez-Riestra et
al. (2004) noted the similarity with SAX~J1819.3--2525/V4641~Sgr which
is also a fast transient that at one time revealed a giant flare of
approximately 10$^{-7}$~\ecs\ (Smith et al. 1999; Hjellming et
al. 2000). Orosz et al. (2001; 2003) and Orosz (2002) determined a
mass between 6.8 and 7.4 $M_\odot$ (1$\sigma$ confidence) from optical
spectroscopy, revealing that it is a black hole (BH) candidate. This
similarity prompts the question whether the accretor in \bron\ is also
a BH.

In this paper we present a 20 ks Chandra observation of \bron. Thanks
to its higher angular resolution and, therefore, smaller background
level, Chandra is able to probe deeper than XMM-Newton which
improves the signal-to-noise ratio of the quiescent emission. Such a
study may yield constraints on the nature of the compact object. If
the luminosity would be above 10$^{32}$~\ecs\ and the spectrum a black
body with a temperature of order a few tens to hundreds of eV, this
would represent strong evidence for a neutron star (NS; Rutledge et
al. 1999; Garcia et al. 2001). Furthermore, the Chandra observations
provide an improvement in the positional accuracy by an order of
magnitude. Thus, the optical counterpart can be identified
unambiguously. This is not the first Chandra observation of a fast
X-ray transient. XTE~J1739--302 was observed for 5 ks (Smith
et al. 2003a, 2003b). However, inspection of those data shows that the
source was not in quiescence.

\begin{figure}[!t]
\centering
\includegraphics[width=0.99\columnwidth]{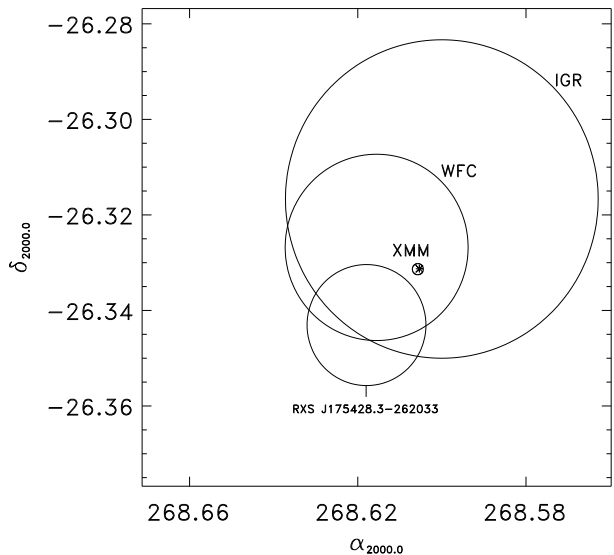}
\caption{Map showing source detections around IGR J17544--2619. All
error circles are for a confidence level in excess of 90\% (the radius
of the 68\%-confident ROSAT error circle was multiplied by two). The
Chandra position is indicated by an asterisk inside the XMM-Newton
error circle. Technically, since the Chandra and XMM-Newton positions
are so close to the ROSAT error circle an association to the ROSAT
cannot be completely ruled out.\label{k1754}}
\end{figure}

\section{Chandra observation}

\begin{figure}[!t]
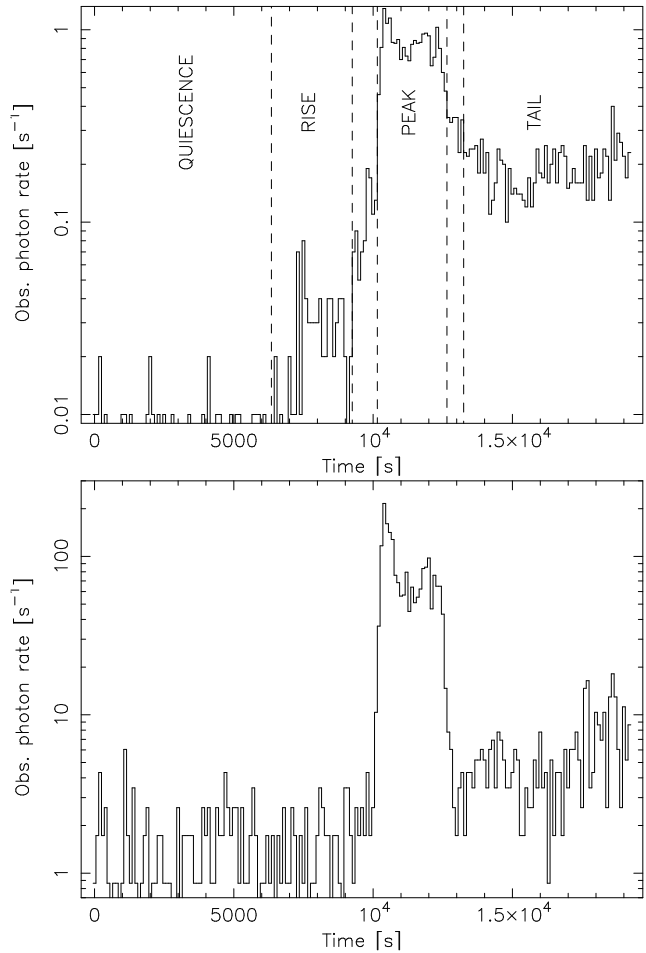

\centering
\includegraphics[height=0.96\columnwidth,angle=270]{Hf291f2a.ps}
\includegraphics[height=0.96\columnwidth,angle=270]{Hf291f2b.ps}
\caption{{\em Top:} Chandra light curve of \bron\ from all photons
extracted from 4\farcs5 from the centroid.  The photon rate is not
corrected for pile up. The background count rate is 2.5$\times10^{-4}$
phot~s$^{-1}$. {\em Bottom:} Light curve from trailed image. No
background subtraction was applied. The photon rate during the first
6300~s, 1.55$\pm0.01$~s$^{-1}$, is due to the
background.\label{chandra_lc}}
\end{figure}

%

\begin{table*}
\begin{center}
\caption[]{Power law spectral fits to the 4 different states indicated
in Fig.~\ref{chandra_lc}. Errors are for 90\%
confidence. \label{chandra_fits}}
\begin{tabular}{lccccc}
\hline\hline
State &  $N_{\rm H}$ (10$^{22}$~cm$^{-2}$) & $\Gamma$ & Obs. flux$^\ddag$ & Unabs. flux$^\ddag$ & $\chi^2_\nu$ \\
\hline
quiescence$^\star$  & 1.36$^\dag$   & 5.9$\pm1.2$   & $ 1.4_{ -0.2}^{ +0.3}\times10^{-14}$ & $ 9.3_{ -5.8}^{+15.4}\times10^{-13}$  & 38\%$^+$ \\
rise$^\star$        & $\downarrow$  & $\downarrow$  & $ 8.4_{ -2.1}^{ +2.1}\times10^{-13}$ & $ 9.8_{ -2.4}^{ +2.4}\times10^{-13}$  & $\downarrow$ \\
peak$^\times$       & $1.36\pm0.22$ & $0.73\pm0.13$ & $2.31_{-0.09}^{+0.09}\times10^{ -9}$ & $2.69_{-0.11}^{+0.11}\times10^{ -9}$  & 0.839 (144 dof) \\
postflare$^\times$  & $\uparrow$    & $\uparrow$    & $1.35_{-0.13}^{+0.13}\times10^{-10}$ & $1.57_{-0.15}^{+0.15}\times10^{-10}$  & $\uparrow$\\
\hline\hline
\end{tabular}
\end{center}

\noindent
$^\dag$Fixed; $^\ddag$Average flux in 0.5--10 keV (\ecs); $^\star$
These spectra were extracted from the point spread function with
extraction radii of 1\farcs5 and 5\farcs0 for the quiescence and rise
data, respectively; $^\times$These spectra were extracted from the
readout trails; $^+$This is the goodness of fit as determined through
10$^4$ Monte Carlo simulations. It is the percentage of simulations
with a goodness of fit parameter better than measured, based on the
best fit model.
\end{table*}

Chandra observed \bron\ on July 3rd, 2004 (obsid 4550), with the
ACIS-S CCD array (Garmire et al. 2001) in the focal plane and no
grating. The CCD frame time is 3.2~s, the exposure time 19.06 ks. We
analyzed these data with CIAO version 3.2.1. The source is clearly
detected. The point source image shows a heavily piled-up source with
a characteristic hole at the center and readout trails. The average
photon position according to the Mexican Hat wavelet method employed
in CIAO tool {\tt wavdetect} is $\alpha_{2000.0}=17^{\rm h}54^{\rm
m}25\fs284$, $\delta_{2000.0}=-26^\circ 19\arcmin52\farcs62$
($l^{II}=3\fdg23$, $b^{II}=+0\fdg33$). The nominal uncertainty is
0\farcs6.  This position is 1\farcs2 from the XMM-Newton position (see
Fig.~\ref{k1754}) and 0\farcs2 from the previously identified optical
counterpart (Rodriguez 2003).

Figure~\ref{chandra_lc} (top) shows the time profile at 100~s
resolution of the rate of all 4191 observed photons within a 4\farcs5
accumulation radius from the centroid. Interestingly, the source shows
two markedly different states. During the first 6300~s the source is
very faint with a mere 26 detected photons. Subsequently, it rises in
two steps over 3700 s to a level that is approximately 250 times
brighter. Thereafter it decays down to a level 4 times fainter at
which it remains for the 5000~s remainder of the observation. To
assess the true dynamic range (corrected for pile up), we made a light
curve of all 2934 photons in the readout trails that are at least
25\arcsec\ from the centroid, employing CIAO tool {\tt
acisreadcorr}. Such counting rates are not subject to pile up for
photon rates below a few thousand s$^{-1}$, because in the readout
mode the CCD pixels have an effective readout time of only
40~$\mu$s. The result is provided in the bottom panel of
Fig.~\ref{chandra_lc}. The peak rate is $214.5\pm2.1$~s$^{-1}$
(background subtracted).  This is 6.5$\times10^4$ times larger than
during quiescence.

We used the 40~$\mu$s resolved trail data to search for a ms pulsar
signal by constructing Fourier spectra for every CCD frame with a net
exposure time of 41~ms, and averaging the spectra. We followed this
procedure for all data as well as for only those pertaining to the
peak and postflare intervals.  No periodic signal was found. The
3$\sigma$ upper limit on the amplitude of a sinusoidal signal is
$\sim$10\%.

We extracted spectra for four characteristic flux levels which are
indicated by dashed lines in Fig.~\ref{chandra_lc} (top): quiescence,
rise, peak and tail. The first two spectra were taken from within
circles centered on the source position with different accumulation
radii. The extraction radius for the quiescent spectrum was chosen to
be 1\farcs5, which is small in order to minimize the background (0.7
background photons are expected in this region). For a soft quiescent
spectrum (see below) the encircled energy is still 90 to 95\% for such
a radius. The spectra for the latter two time intervals, where pile up
is heavy, were taken from the readout trails.  Quick inspection of the
spectra reveals that there is an obvious change in the hardness of the
spectrum between quiescence and thereafter. All photons of the former
spectrum are detected between 1.0 and 2.2 keV while for the rising
part and the peak part the fraction of photons above 2.2 keV is 72\%
and 74\%, respectively.

We modeled all spectra with an absorbed power law, using XSPEC
vs. 11.3.1 (Arnaud et al. 1996). For the quiescent spectrum we applied
the Cash statistic (Cash 1979) because of the low number of photons;
for the remaining spectra the chi-squared statistic was used, after
rebinning energy channels so that each channel contains at least 15
photons.  The three non-quiescence spectra were fitted simultaneously,
employing a single free hydrogen column density, which parametrizes
the low-energy absorption (Morrison \& McCammon 1983), and a single
photon index $\Gamma$. The results are provided in
Table~\ref{chandra_fits}. The fitted value for $N_{\rm H}$ is, within
errors, identical to the interstellar value of
$(1.4\pm0.2)\times10^{22}$~H-atoms cm$^{-2}$ (Dickey \& Lockman
1990). When leaving free $N_{\rm H}$ or $\Gamma$, the fit did not
improve (i.e., $\chi^2_\nu$ did not decrease).

The quiescent spectrum could also be satisfactorily fitted with a NS
hydrogen atmosphere model following Pavlov et al. (1992) and Zavlin
(1996), with $N_{\rm H}$ fixed to $1.36\times10^{22}$~H-atoms
cm$^{-2}$. The fit was similar in quality: 37\% of 10$^4$ Monte Carlo
simulations based on the best-fit model had a smaller C statistic than
observed.  We find for a NS at 3.5 kpc with $M=1.4~M_\odot$, $R=$10~km
and $B$=0~G a temperature of $90\pm10$~eV. The unabsorbed 0.5--10 keV
flux is $(1.9\pm0.4)\times10^{-13}$~\ecs.

\section{Discussion}

The observation was fortunately timed catching the source both in
quiescence and outburst. The outburst phase provides an unambiguous
identification and the quiescence phase a constraint for the nature of
the accretor as follows. Kong et al. (2002) reports Chandra ACIS-S
spectra of four quiescent BH systems and find that these can all be
best modeled with an (absorbed) power law with $\Gamma$ between
1.70$^{+0.88}_{-0.78}$ and 2.28$^{+0.47}_{-0.64}$ (90\% confidence
level error margins).  Other simple models also fit the data, except
the black body model. The power law model is consistent with an
advection-dominated accretion flow according to Kong et al. Similar
results were obtained on other BH candidates (Tomsick et
al. 2003). The quiescent spectrum of \bron\ does not adhere to this
rule. That suggests that \bron\ is {\em not} a BH candidate, but a
NS. The quiescent flux corrected for absorption and extrapolated to
0.01--10 keV is $(3.5\pm0.9)\times10^{-13}$~\ecs, based on the NS
model spectrum.  This translates to a luminosity at 3.5~kpc of
$(5.2\pm1.3)\times10^{32}$~\lum, a reasonable value for a NS.

However, OB stars also emit X-rays with similarly soft spectra. In a
list of optically bright OB-type stars observed with ROSAT
(Berghh\"{o}fer et al. 1997a and 1997b) there are 14 O9I/II
supergiants. Six were detected with luminosities between 1.7 and
$7.1\times10^{32}$~\lum\ and eight were not detected with upper limits
from 0.4 to $1.7\times10^{32}$~\lum. Thus, a considerable portion of
the observed luminosity may be explained by radiation from the donor
star. Given the statistics, though, it is more likely that most comes
from the accretor.

We speculate on the origin of the outbursts, that are so brief
compared to typical transient X-ray binary outbursts of at least a
week.  Our study shows that this is probably unrelated to the nature
of the compact object, given that \bron\ probably contains a NS while
V4641~Sgr contains a BH (Orosz et al. 2001). Rather, it must be
related to the donor star.  As already noted by In~'t~Zand et
al. (2004) and Smith (2004), optical counterparts for three fast X-ray
transients have been identified and all are of early spectral type
(O9Ib for \bron, O8 for XTE~J1739--302 and B9 for V4641~Sgr). Many
early-type stars are suspected to have massive winds driven by
resonance line scattering that are highly structured and variable
(e.g., review by Owocki 1997; see also Lamers \& Cassinelli 1999). A
star that is particularly interesting in the present context is the B0
main sequence star $\tau$~Sco whose otherwise homogeneous wind is
thought to contain at any time $10^3$ clumps with masses of order
10$^{19-20}$~g (Howk et al. 2000). A sizeable fraction of these may
fall back to the star, giving rise to shock X-ray emission and
redshifted spectral features, while others may flow with the
wind. Thus, it is fairly well established that clumpy winds can exist
in early-type stars and clump capture by a nearby compact object may
explain the time scale of the X-ray flares. Surprisingly, the masses
of the clumps in $\tau$ Sco are a good match to the fluence of the
transient outburst within an order of magnitude, assuming that all
gravitational energy provided by the fall of the clump on a NS is
liberated in the form of radiation.  Modeling of this effect for a
supergiant rather than main sequence star and for an earlier spectral
class than for $\tau$~Sco should be carried out to verify whether the
outburst duration and recurrence time make sense in this scenario.

Our improved positional accuracy (by a factor of 6 in one axis)
confirms the identification with \optical. An initial spectroscopic
study of that star is being carried out by Pellizza et al. (in
prep.). As a follow-up it is important to carry out time-resolved
spectroscopy of this star. Measurements of Doppler-shifted spectral
lines may provide constraints on key parameters such as the orbital
period and the mass of the compact object. The shape of the lines is
instrumental in determining the outburst mechanism: it may show signs
of the clumpiness of the wind (e.g., Howk et al. 2000) or of strong
variability in wind speed and/or density (e.g., Lindstr{\o}m et
al. 2005). Preferably, there should be simultaneous optical and X-ray
observations.

\acknowledgement 

R. Remillard is thanked for checking out high RXTE/ASM data points,
P. Jonker, H. Lamers, M. M\'{e}ndez and T. Raassen for useful
discussions, and S. Chaty for sharing preliminary results on an
optical study.  JZ acknowledges support from the Netherlands
Organization for Scientific Research (NWO).

\end{document}